\documentclass[twocolumn,twoside,slac_two]{revtex4}
\usepackage{graphicx}
\usepackage{fancyhdr}
\newcommand{\gray}{$\gamma$-ray\ } \newcommand{\grays}{$\gamma$-rays\ }
 
\begin{document}

\title{An Empirical Determination of the EBL and the Gamma-ray Opacity of the Universe}

%

\author{Floyd W. Stecker}
\affiliation{NASA Goddard Space Flight Center, Greenbelt, MD 20771, USA}
\affiliation{Department of Physics and Astronomy, University of California, Los Angeles}
%

\begin{abstract}

I present the results of a new approach to the intensity and photon density spectrum of the intergalactic background light as
a function of redshift using observational data obtained in many different
wavelength bands from local to deep galaxy surveys. This enables an empirical determination of both the EBL and its observationally based uncertainties. Using these results one can place 68\% confidence upper and lower limits on the opacity of the universe to $\gamma$-rays, free of
the theoretical assumptions that were needed for past calculations. I compare
our results with measurements of the extragalactic background light, upper limits obtained
from observations made by the {\it Fermi} ~Gamma-ray Space Telescope, and new observationally based results from {\it Fermi} and {\it H.E.S.S.} using recent analyses of blazar spectra.

\end{abstract}

\maketitle

\thispagestyle{fancy}


\section{INTRODUCTION}
Past work on estimating the spectral and redshift characteristics of the intergalactic 
photon density, generically referred to as the EBL, have depended on various assumptions as to
the evolution of stellar populations and dust absorption in galaxies. A detailed review
of the problem has been given by Dwek \& Krennrich \cite{dwe12}. There have also been attempts to
probe the EBL from studies of blazar spectra \cite{ack12, abr12} an approach originally suggested by Stecker, De Jager \& Salamon \cite{sds92}. In this paper, I 
present the results of a new, fully empirical approach to calculating the EBL, and subsequently, the $\gamma$-ray opacity of the Universe from pair production interactions of $\gamma$-rays with the EBL. This approach of M.A. Malkan, S.T. Scully and myself, hitherto unattainable, is now enabled by very recent data from deep galaxy surveys spanning the electromagnetic spectrum from millimeter to UV wavelengths and therewith using galaxy luminosity functions for redshifts up to $z = 8$. This new approach, in addition to being an alternative 
to the approaches mentioned above, is {\it totally model independent}; it does not make assumptions as to galaxy dust or star formation characteristics or as to blazar emission models. It is also an approach uniquely capable of {\it delineating empirically based uncertainties on the determination of the EBL}. More details of the work presented here have now been published \cite{sms12}.(See also Ref. \cite{hk12}.)

\section{GALAXY EMISSIVITIES AND LUMINOSITY DENSITIES}
The observationally determined co-moving radiation energy density $u_{\nu}(z)$, is
derived from the co-moving specific emissivity ${\cal E}_{\nu}(z)$,
which, in turn is derived from the observed galaxy luminosity function (LF) at redshift $z$. 
The galaxy luminosity function,~$\Phi_{\nu}(L)$, is defined
as the distribution function of galaxy luminosities at a specific frequency
or wavelength. The specific emissivity at frequency $\nu$ and redshift $z$ 
(also referred to in the literature as the
luminosity density, $\rho_{L_{\nu}}$) , is the integral over the luminosity function

\begin{equation}
 {\cal E}_{\nu}(z) = \int_{L_{min}}^{L_{max}} dL_{\nu} \, \Phi(L_{\nu};z)
\label{phi}
\end{equation}  

In view of the well known difficulties in predicting galaxy LFs based on galaxy formation
models \cite{mar07}, our working philosophy was to depend only on observational determinations of galaxy LFs in deriving specific emissivities.

There are many references in the literature where the LF is given and fit to Schechter
parameters, but where $\rho_{L_{\nu}}$ is not given. In those cases, we could not determine
the covariance of the errors in the Schechter parameters used to determine the dominant statistical errors in their analyses. Thus, we could not ourselves accurately determine
the error on the emissivity from equation (\ref{phi}). We therefore chose to use only the
papers that gave values for $\rho_{L_{\nu}}(z) = {\cal E}_{\nu}(z)$ with errors. We did not consider cosmic variance, but this uncertainly should be minimized since we used data from many surveys.

The co-moving radiation energy density $u_{\nu}(z)$ 
is the time integral of the co-moving specific emissivity ${\cal E}_{\nu}(z)$,
\begin{equation} 
\label{u1}
u_{\nu}(z)=
\int_{z}^{z_{\rm max}}dz^{\prime}\,{\cal E}_{\nu^{\prime}}(z^{\prime})
\frac{dt}{dz}(z^{\prime})e^{-\tau_{\rm eff}(\nu,z,z^{\prime})},
\end{equation}

\noindent where $\nu^{\prime}=\nu(1+z^{\prime})/(1+z)$ and $z_{\rm max}$ is the
redshift corresponding to initial galaxy formation 
\cite{sal98}, and
\begin{equation}
\frac{dt}{dz}{(z)} = {[H_{0}(1+z)\sqrt{\Omega_{\Lambda} + \Omega_{m}(1+z)^3}}]^{-1},
\label{cosmology}
\end{equation}

\noindent with $\Omega_{\Lambda} = 0.72$ and $\Omega_{m} = 0.28$.

The opacity factor for frequencies below the Lyman limit is dominated by dust extinction.
Since we were using actual observations of galaxies rather than models, dust absorption is implicitly included. The remaining opacity $\tau_{\nu}$ refers to the extinction of ionizing photons
with frequencies above the rest frame Lyman limit of $\nu_{LyL} \equiv 3.29 \times 10^{15}$ Hz by interstellar and intergalactic hydrogen and helium. It has been shown that this opacity is 
very high, corresponding to the expectation of very small fraction of ionizing radiation in
intergalactic space compared with radiation below the Lyman limit \cite{lyt95, sal98}. In fact, the Lyman limit cutoff is used as a tool; when galaxies disappear when
using a filter at a given waveband ({\it e.g.}, "$U$-dropouts", "$V$-dropouts") it is an indication of the redshift of the Lyman limit. 

We have therefore replaced equation (\ref{u1}) with the following expression

\begin{equation} 
\label{u2}
u_{\nu}(z)=
\int_{z}^{z_{\rm max}}dz^{\prime}\,{\cal E}_{\nu^{\prime}}(z^{\prime})
\frac{dt}{dz}(z^{\prime}){{\cal H}(\nu(z') - \nu'_{LyL})},
\end{equation}

\noindent where ${\cal H}(x) $ is the Heavyside step function.

\section{EMPIRICAL SPECIFIC EMISSIVITIES}
We have used the results of many galaxy surveys to compile a set of luminosity
densities (LDs), $\rho_{L_{\nu}}(z) = {\cal E}_{\nu}(z)$, at all observed redshifts, and at rest-frame wavelengths from the far-ultraviolet, FUV = 150 nm  to the $I$ band, $I$ = 800 nm.  
The LDs were obtained with a wide variety of instruments in many different deep fields \cite{sms12}.
Figure 1 shows the redshift evolution of the luminosity ${\cal E}_{\nu}(z)$ for the various wavebands based on those published in the literature. In order to determine the redshift evolution of the LD in each of the bands out to a redshift of $\sim$ 8 where only UV data are available, we utilized observed color relations to transform data from other bands. We have chosen to include all data possible at $z > 1.5$ in order to to fill in the observational gaps for various wavebands, mostly at higher redshifts. We used the redshift-dependent observations of average galaxy colors where appropriate in our analysis. In the redshift ranges where they overlap, {\it the colored (observational) data points shown in Figure 1 for the various wavelength bands agree quite well (within the uncertainties) with the black data points that were extrapolated from the shorter wavelength bands using our color relations.} The observationally determined LDs, combined with the color relations, extend our coverage of galaxy photon production from the FUV to NIR wavelengths in the galaxy rest frame. Our final results are not very sensitive to errors in our average color relations because the interpolations that we made were only over very small fractional wavelength intervals, $\Delta\lambda(z)$. We have directly tested this by using numerical trial runs.

\begin{figure*}[t]
\centering
\includegraphics[height=7in]{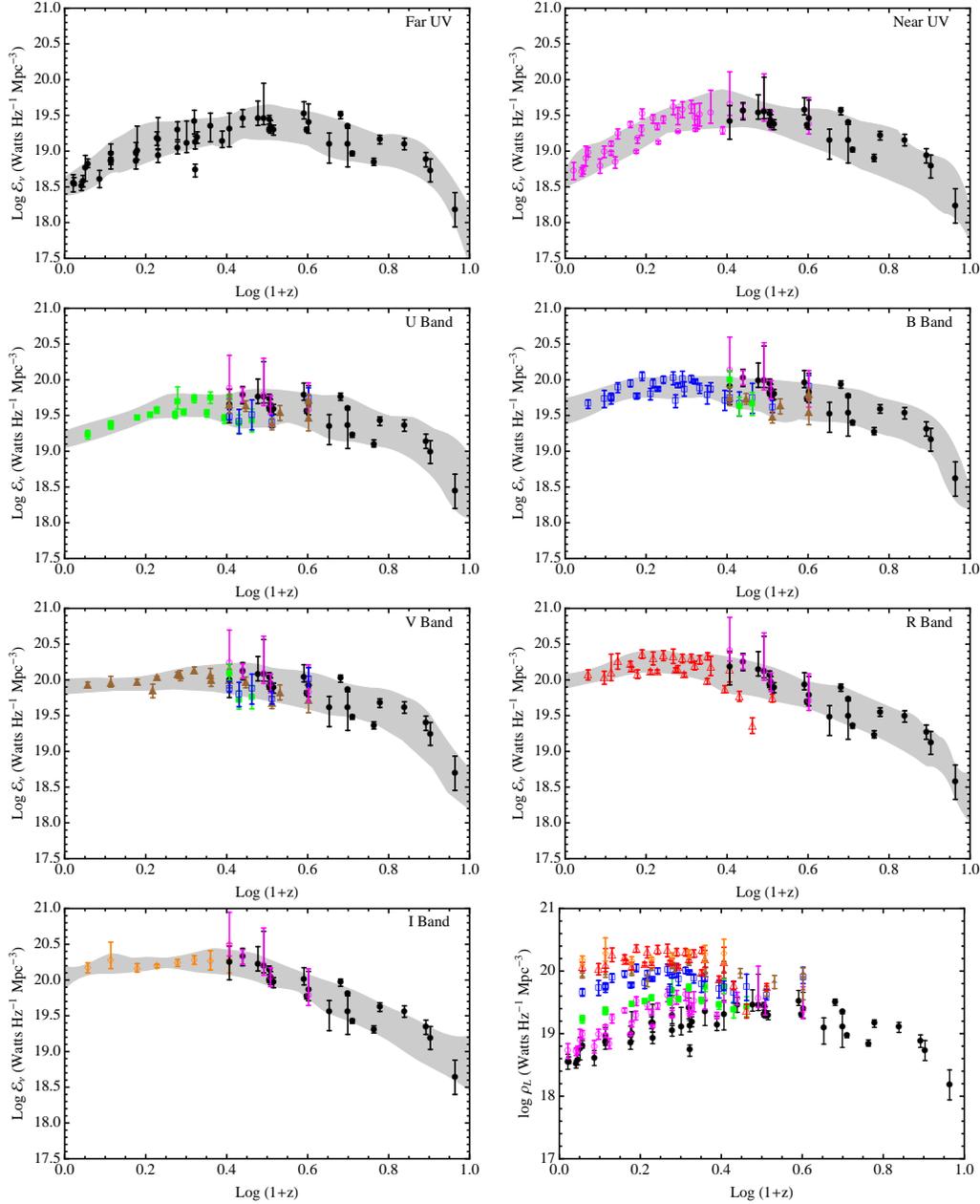}
\caption{The observed specific emissivities in our standard astronomical fiducial wavebands.  The lower right panel shows all of the observational data used. 
In the other panels, non-band data have been shifted using observed color relations
in order to fully determine the specific emissivities in each waveband. The symbol designations are FUV: black filled circles, NUV: magenta open circles, $U$: green filled squares, $B$: blue open squares, $V$: brown filled triangles, $R$: orange open triangles, 
$I$: yellow open diamonds.  Grey shading: derived 68\% confidence bands.} 
\label{pconf}
\end{figure*}

\section{THE PHOTON DENSITIES WITH EMPIRICAL UNCERTAINTIES}
The 68\% confidence band upper and lower limits of the EBL were determined from the observational data on $\rho_{L_{\nu}}$. We made no assumptions about luminosity density evolution.  We derived a luminosity confidence band in each waveband by using a robust rational fitting function characterized by
\begin{equation}
\rho_{L_{\nu}} = {\cal E}_{\nu}(z) = {{ax+b}\over{cx^{2}+dx+e}}
\label{rational}
\end{equation}
where $x = \log(1+z)$ and $a$,$b$,$c$,$d$,and $e$ are free parameters.

The 68\% confidence band was then computed
from Monte Carlo simulation by finding $10^5$ realizations of the data and then fitting the  function to the form given by eq. (\ref{rational}). In order to best represent the tolerated confidence band, particularly at the highest redshifts, we chose to equally weight all FUV points in excess of a redshift of 2.  {\it Our goal was not to find the best fit to the data, but rather to find the limits tolerated by the current observational data.}  In order to perform the Monte Carlo analysis of the fitting function, a likelihood was determined at each redshift given the existing data.  The shape of this function was taken to be Gaussian (or the sum of Gaussians where multiple points exist) for symmetric errors quoted in the literature.  Where symmetric errors are not quoted it is impossible to know what the actual shape of the likelihood function is.  We have chosen to utilize a skew normal distribution to model asymmetric errors.   This assumption has very little impact on the determination of the confidence bands. The resulting bands are shown along with the luminosity density data in Figure 1.

With the confidence bands established, we took the upper and lower limits of the bands to be our high and low EBL constraints respectively.  We then interpolated each of these cases separately  between the various wavebands to find the upper and lower limit rest frame LDs.  The calculation was extended to the Lyman limit using the derivative derived from our color relationship between the near and far UV bands.  The co-moving radiation energy density was then determined from equation (\ref{u2}). This result was used as input for the determination of the optical depth of the universe to \grays . Our resultant $z = 0$ EBL is shown in Figure 2 and compared with the present data, as discussed in Ref. \cite{sms12}.

\begin{figure*}[t]
\centering
\includegraphics[height=3.35in]{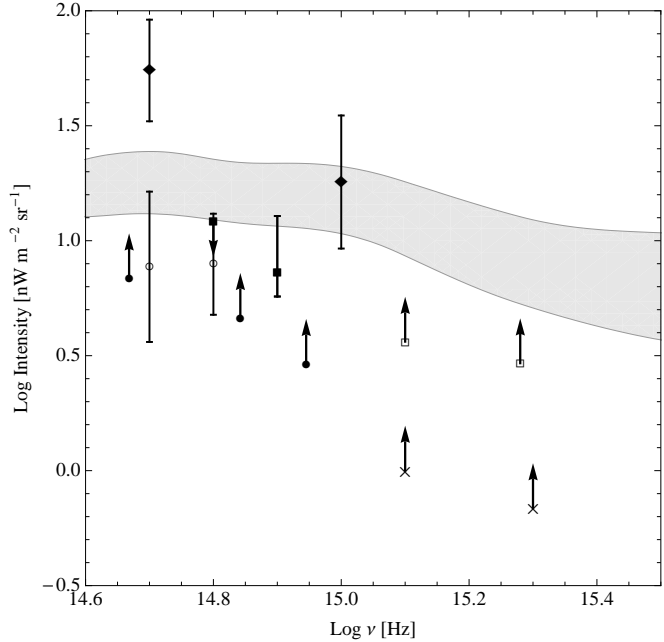}
\caption{Our empirically-based determination of the EBL together with lower limits and data as described in Ref. \protect\cite{sms12}.} 
\label{EBL}
\end{figure*}

\section{THE OPTICAL DEPTH FROM $\gamma-\gamma$ INTERACTIONS WiTH UV-IR PHOTONS}

The photon density
\begin{equation}
n(\epsilon,z) = u(\epsilon,z)/\epsilon \ \ ,
\label{gammadens}
\end{equation}
with $\epsilon = h\nu$, $h$ being Planck's constant, were
calculated using equation (\ref{u1}).

The cross section for photon-photon scattering to electron-positron pairs can be calculated using quantum electrodynamics \cite{bre34}.  
The threshold for this interaction is determined from the frame invariance of the square of the four-momentum vector that reduces to the square of the threshold energy, $s$, required to produce twice the electron rest mass in the c.m.s.,  
\begin{equation}
s = 2\epsilon E_{\gamma} (1-\cos\theta) = 4m_{e}^2
\label{s}
\end{equation}

This invariance is known to hold to within one part in $10^{15}$ \cite{ste01, jac04}.
With the co-moving energy density $u_{\nu}(z)$ evaluated, the optical
depth for \grays owing to electron-positron pair production 
interactions with photons of the stellar radiation
background can be determined from the expression \cite{sds92}
\begin{eqnarray*} 
\label{tau}
\tau(E_{0},z_{e})  & = & c\int_{0}^{z_{e}}dz\,\frac{dt}{dz}\int_{0}^{2}
dx\,\frac{x}{2}\int_{0}^{\infty}d\nu\,(1+z)^{3}
 \\ 
& & \times  \ \left[\frac{u_{\nu}(z)}{h\nu}\right] \sigma_{\gamma\gamma}[s=2E_{0}h\nu x(1+z)],
\label{tau}
\end{eqnarray*}

\noindent where $E_{0}$ is the observed \gray energy at redshift zero, 
$\nu$ is the frequency at
redshift $z$,
$z_{e}$ is the redshift of
the \gray source at emission, $x=(1-\cos\theta)$, \\
$\theta$ being the angle between the \gray and the soft background photon.

The pair production cross section $\sigma_{\gamma\gamma}$ is zero for
center-of-mass energy $\sqrt{s} < 2m_{e}c^{2}$, $m_{e}$ being the electron
mass.  Above this threshold, the pair production cross section is given by

\begin{eqnarray*} 
\label{sigma}
\sigma_{\gamma\gamma}(s) & = & \frac{3}{16}\sigma_{\rm T}(1-\beta^{2}) \\
& & \times \ \left[ 2\beta(\beta^{2}-2)+(3-\beta^{4})\ln\left(\frac{1+\beta}{1-\beta}
\right)\right],
\end{eqnarray*} 

\noindent where $\sigma_T$ is the Thompson scattering cross section and $\beta=(1-4m_{e}^{2}c^{4}/s)^{1/2}$ \cite{jau55}.

It follows from equation (\ref{s}) that the pair-production cross section energy has a threshold at $\lambda = 4.75 \ \mu {\rm m} \cdot E_{\gamma}({\rm TeV})$. 
\noindent Since the maximum $\lambda$ that we consider
here is in the rest frame I band at 800 nm at redshift $z$, and we observe $E_{\gamma}$ at redshift 0, so that its energy at
interaction in the rest frame is $(1+z)E_{\gamma}$, we then get a
conservative upper limit on $E_{\gamma}$ of $\sim 200(1+z)^{-1}$ GeV as the maximum \gray energy
affected by the photon range considered here. Allowing for a small error, our opacities
are good to $\sim 250(1+z)^{-1}$ GeV.
The 68\% opacity ranges for $z = 0.1,0.5, 1, 3 ~$and $5$ are plotted in Figure 3. 

The widths of the grey uncertainty ranges in the LDs shown in Figure 1 increase towards higher redshifts, especially at the longest rest wavelengths.  This reflects
the decreasing amount of long-wavelength data and the corresponding increase in
uncertainties about the galaxies in those regimes. However,
these uncertainties do not greatly influence
the opacity calculations. Because of the short time interval of the emission from
galaxies at high redshifts their photons do not contribute greatly to the opacity at lower redshifts. Figure 3 shows that the opacities determined for redshifts of 3 and 5 overlap
within the uncertainties.

\begin{figure}
\begin{center}
\includegraphics[height=3.35in]{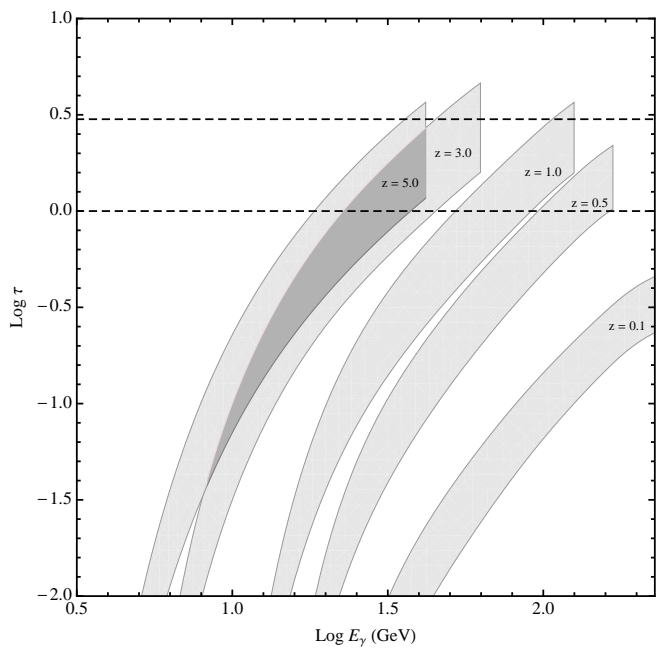}
\label{opacities}
\end{center}
\vspace{-0.3in}
\caption{The empirically determined opacities for redshifts of 0.1, 0.5, 1, 3, 5.
The dashed lines are for $\tau = 1$ and $\tau = 3$ \protect\cite{sms12}.}
\end{figure}

\section{Results and Implications }

We have determined the EBL using local and deep galaxy survey data, together with observationally produced uncertainties, for wavelengths from 150 nm to 800 nm and redshifts out to $z > 5$. We have presented our results in terms of  68\% confidence band upper and lower limits. 
In Figure 2, we compare our $z = 0$ result with both published and preliminary measurements and limits. As expected, our $z = 0$ (EBL) 68\% lower limits as shown in Figure 2 are higher than those obtained by galaxy counts alone, since the EBL from galaxies is not completely resolved.

Figure 4 shows our 68\% confidence band for $\tau = 1$ on an energy-redshift plot \cite{fs70}
compared with the {\it Fermi} data on the highest energy photons from extragalactic
sources at various redshifts as given in Ref. \cite{abd10}. It can be seen that none of the
photons from these sources would be expected to be significantly annihilated by pair
production interactions with the EBL. This point is brought out further in Figure 5.
This figure compares the 68\% confidence band of our opacity results with the 95\% confidence
upper limits on the opacity derived for specific blazars \cite{abd10}.

In a recent publication, the {\it Fermi} Collaboration has probed for the imprint of the intergalactic radiation fields (the EBL) in the \gray spectra of blazars \cite{ack12}, an approach originally suggested in Ref. \cite{sds92}. Their result appears to be consistent with our results
near the low opacity end of our uncertainty range. The {\it H.E.S.S.} group \cite{abr12} has also looked for such an effect in the spectra of bright blazars at energies above 100 GeV. It follows from eq. (\ref{s}) that such air \v{C}erenkov telescope studies are sensitive only to interactions of \grays with infrared photons. The {\it H.E.S.S.} group has recently obtained a value for the $z = 0$ EBL of $15 \pm 2_{stat} \pm 3_{sys}$ nW m$^{-2}$sr$^{-1}$ at a wavelength of 1.4 $\mu$m. 
This compares to our value of $17.5 \pm 4.9$ nW m$^{-2}$sr$^{-1}$ at 0.9 $\mu$m.

My colleagues M.A. Malkan and S.T. Scully and I are presently continuing our studies of the
photon density spectrum as a function of redshift into the infrared range using surveys from 
{\it Hubble}, {\it Spitzer}, {\it Herschel} and other sources. Such studies, together with ongoing complementary \gray observations of extragalactic sources with {\it Fermi} and future observations using the {\it \v{C}erenkov Telescope Array}, which will be sensitive to energies above 10 GeV 
\cite{cta10}, one can look forward to a obtaining better understanding of both the EBL and other potential aspects of \gray propagation in the Universe, such as those explored in Refs. \cite{dea09} -- \cite{ess12}.

\begin{figure}
\begin{center}
\includegraphics[width=3.25in]{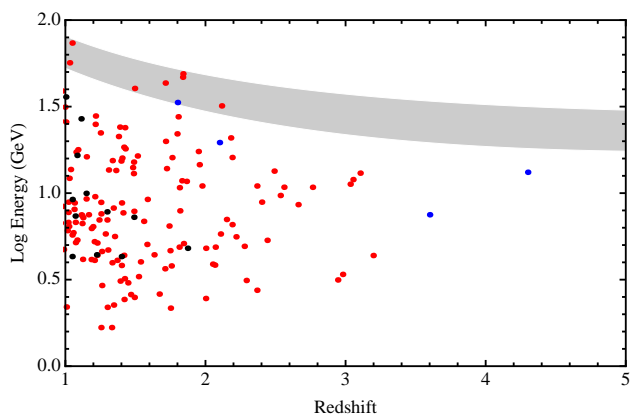}
\label{FSplot}
\end{center}
\vspace{-0.3in}
\caption{An energy-redshift plot of the \gray horizon 
showing our uncertainty band results \protect\cite{sms12} compared with the {\it Fermi} plot of their highest energy photons from FSRQs (red), BL Lacs (black) and GRBs (blue)  vs. redshift
\protect\cite{abd10}.}
\end{figure}

\begin{figure*}
\begin{center}
\includegraphics[width=7.0in]{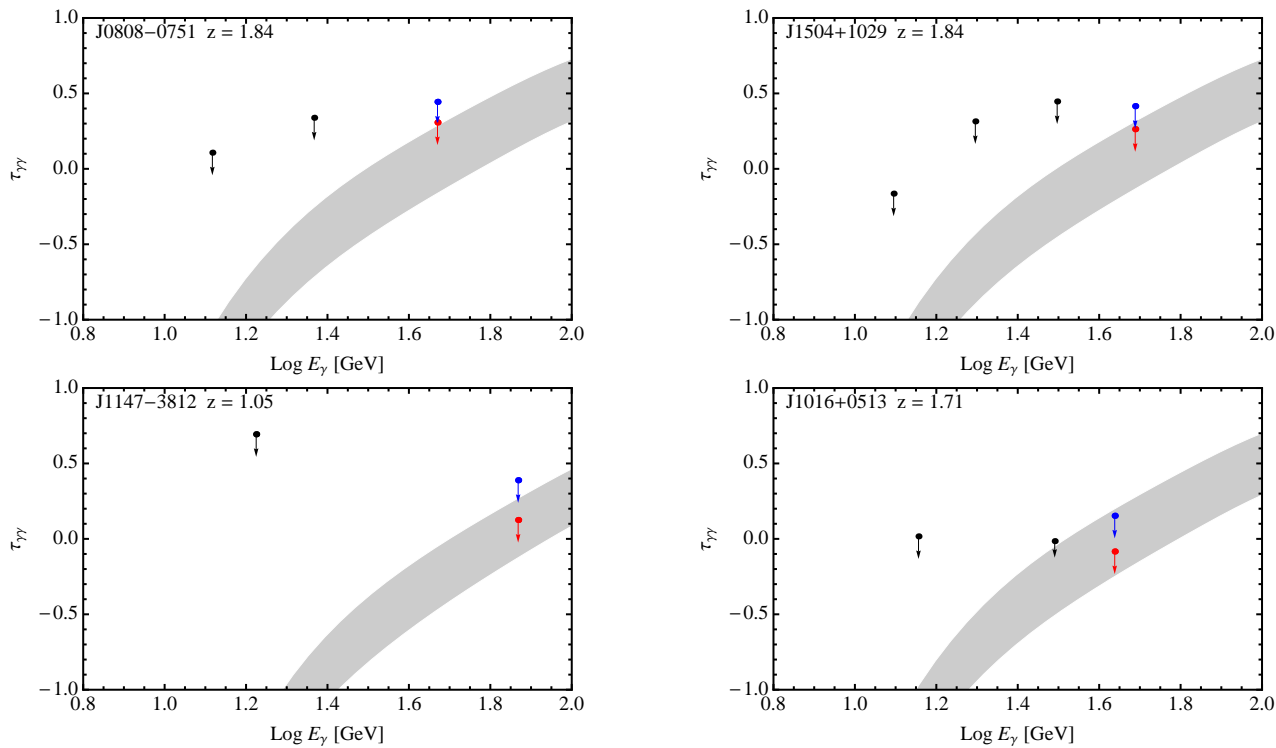}
\label{Fermilim}
\end{center}
\vspace{-0.4in}
\caption{Our opacity results for the redshifts of the blazars indicated \protect\cite{sms12} compared with \protect95\% confidence opacity upper limits (red arrows) and \protect99\% confidence limits (blue arrows) as given by the {\it Fermi} analysis \protect\cite{abd10}.}
\end{figure*}

\subsection*{Results Online}

Our results in numerical form are available at the following link:

\noindent {\tt http://csma31.csm.jmu.edu/physics/} \\
 {\tt scully/opacities.html}

\begin{acknowledgments}
F.W.S., M.A. Malkan and S.T. Scully were partially supported by a Fermi Cycle 4 Guest Investigator grant.
\end{acknowledgments}



\end{document}